     \documentstyle[12pt,epsfig]{article}
     \newlength{\dinwidth}                       
     \newlength{\dinmargin}                      
\def\lsim{\mathrel{\rlap{\lower4pt\hbox{\hskip1pt$\sim$}}
    \raise1pt\hbox{$<$}}}                
\def\gsim{\mathrel{\rlap{\lower4pt\hbox{\hskip1pt$\sim$}}
    \raise1pt\hbox{$>$}}}                
\newcommand{\fb }{{\rm fb }}
\newcommand{\ab }{{\rm ab }}
\newcommand{\TeV}{{\rm TeV}}
\newcommand{\GeV}{{\rm GeV}}

\begin{document}
\thispagestyle{empty}
\begin{flushleft}
September 2000 \hfill {\tt hep-ph/0009323} \\
{\sf LC-TH-2000-050}
\end{flushleft}

\setcounter{page}{0}

\mbox{}
\vspace*{\fill}
\begin{center}{\Large \bf
Precision of electro-weak couplings of\\

\vspace{2mm}
scalar leptoquarks at TESLA
\footnotemark}
\footnotetext{Contribution to the TESLA TDR. 
                            }

\vspace{5em}
\large
Johannes Bl\"umlein
\\
\vspace{5em}
\normalsize
{\it     Deutsches Elektronen-Synchrotron, DESY\\
Platanenallee 6, D-15738 Zeuthen, Germany}\\

\vspace*{\fill}
\end{center}
\begin{abstract}
\noindent
We investigate the potential to measure the electro-weak couplings of
scalar leptoquarks $\Phi_s$ at TESLA for energies in the range of
$\sqrt{s} \simeq 1 \TeV$ using the pair production process
$e^+e^- \rightarrow \Phi_s \overline{\Phi}_s$.
\end{abstract}

\vspace{1mm}
\noindent

\vspace*{\fill}


\newpage
\vspace*{10mm}
\begin{center}  \begin{Large} \begin{bf}
Precision of electro-weak couplings of
scalar leptoquarks at TESLA
\\
  \end{bf}  \end{Large}
  \vspace*{5mm}
  \begin{large}
Johannes  Bl\"umlein                               \\
  \end{large}

\vspace{2mm}
         Deutsches~Elektronen-Synchrotron~DESY,
     Platanenallee 6,~D-15738~Zeuthen,~Germany\\

\end{center}
%
\begin{quotation}
\noindent
{\bf Abstract:}
We investigate the potential to measure the electro-weak couplings of
scalar leptoquarks $\Phi_s$ at TESLA for energies in the range of
$\sqrt{s} \simeq 1 \TeV$ using the pair production process
$e^+e^- \rightarrow \Phi_s \overline{\Phi}_s$.
\end{quotation}
%

\vspace{1mm}
\noindent
Leptoquarks are hypothetical particles which combine quantum numbers
of the fundamental fermions of the Standard Model and emerge as bosonic
(scalar and vector) states in various extensions of the Standard Model 
such as unified theories and sub-structure models. In most of the 
scenarios the mass spectrum of these states is not predicted. In a series
of models, however, one expects states in the range of several hundred 
GeV to a few TeV. As being colour (anti)triplets these particles are
likely to be found first at TEVATRON or LHC in strong interaction
processes~\cite{HAD}. There the
corresponding cross sections are widely 
independent of the $SU(2)_L \otimes U(1)_Y$ gauge charges of these 
particles which, on the other hand,
can ideally be determined via the pair production
process $e^+ e^- \rightarrow \Phi \overline{\Phi}$~\cite{EW}
 at high energy linear
colliders. Since the fermionic couplings $\lambda_{lq}$
of leptoquarks are found to be rather small~\cite{lqLQ} the production 
process is completely determined by the electroweak couplings of these 
particles for scalar leptoquarks~\cite{EW} and additional anomalous
couplings for vector leptoquarks (cf.~\cite{VECee} for details).
Currently the mass range of $M_{\Phi} \lsim 250 \GeV$ is excluded for
the 1st family  leptoquark states~\cite{LQ1}, and the mass limits
for 2nd and 3rd family leptoquarks are 200 and 100 GeV, respectively,
\cite{LQ2}. Searches at TESLA will concentrate
on the mass range up to $M_\Phi \lsim 500 \GeV$ at cms energies
$\sqrt{s} \lsim 1 \TeV$ assuming an integrated luminosity of ${\cal L}
=   1~{\rm ab}^{-1}$. The production cross section for scalar
leptoquarks is given by~\cite{EW}
\begin{equation}
\sigma_s = \frac{\pi \alpha^2 \beta^3}{2s} \sum_{a =L,R}
\left| \sum_{V = \gamma, Z} Q_a^V(e) \frac{s}{s - M_V^2 + i M_V \Gamma_V}
Q^V_{\Phi}\right |^2
\end{equation}
with $Q_{L,R}^\gamma = -1, 
Q_L^Z = (-1/2 + \sin^2(\theta_w))/\cos(\theta_w)\sin(\theta_w),
Q_R^Z = \tan(\theta_w), Q^\gamma_\Phi = Q_{em}, Q^Z_\Phi = (T_3 -
Q_{em} \sin^2(\theta_w))/\cos(\theta_w)\sin(\theta_w)$, where $\theta_w$
is the weak mixing angle, $T_3$ is $z$-component of the weak isospin
of the leptoquark, $Q_{em}$ its electric charge  and 
$\beta = \sqrt{1- 4M^2_\Phi/s}$. $M_V$ and $\Gamma_V$ denote mass and
width of the gauge bosons $\gamma$ and $Z$. 
The Beamstrahlung and QED initial state radiation contribution to
$O(\alpha^2)$ do widely cancel against the QED and QCD final state
radiation contributions, see.~Ref.~\cite{RC}. Therefore the Born cross
section yields already a reasonable first estimate for the production
cross section.
Following the notation for
the leptoquarks \cite{BRW} the statistical precisions of the electro-weak
charges $\delta Q_\Phi^{\gamma,Z}$ are summarized in table~1 considering
the examples of leptoquark pair production at 
$\sqrt{s} = 800 \TeV$,
$M_\Phi = 320 \GeV$ and ${\cal L} = 500~{\rm fb}^{-1}$ (1st line) and
$\sqrt{s} = 1 \TeV$,
$M_\Phi = 400 \GeV$ and ${\cal L} = 1~{\rm ab}^{-1}$ (2nd line).
The  statistical accuracies to be obtained are similar in both cases
and range between $\pm 0.004$ and $\pm 0.009$
for $Q_\Phi^\gamma$ and between $\pm 0.008$ and $\pm 0.04$ for
$Q_\Phi^Z$.
\renewcommand{\arraystretch}{1.4}
\begin{center}
\begin{tabular}{||c||r|r|r|r|c|c||       }
\hline \hline
\multicolumn{1}{||c||}{$\Phi_s$}          &
\multicolumn{1}{|c}{$Q_{\Phi}^{\gamma}$} &
\multicolumn{1}{|c}{$T_3$}               &
\multicolumn{1}{|c}{$Q_{\Phi}^{Z}$}      &
\multicolumn{1}{|c}{\#}                  &
\multicolumn{1}{|c}{$\delta Q_{\Phi}^{\gamma}$} &
\multicolumn{1}{|c||}{$\delta Q_{\Phi}^{Z}$}          \\
\hline \hline
$S_1$          &  1/3 &  0   &--0.182&  1419    &$\pm 0.005$
&$+0.035-0.028$ \\
               &      &      &       &  1815    &$\pm 0.004$
&$+0.031-0.025$ \\
\hline
$\tilde{S}_1$  &  4/3 &  0   &--0.729&  22706   &$\pm 0.005$
    & $\pm 0.032$    \\
               &      &  0   &       &  29042   &$\pm 0.004$
    & $\pm 0.028$   \\
\hline
$S_3^u$        &  4/3 &  1   &  1.648& 36020    &$\pm 0.006$
& $\pm 0.012$     \\
               &      &      &       & 45946    &$\pm 0.005$
& $\pm 0.010$     \\
$S_3^0$        &  1/3 &  0   &--0.182&  1419     &$\pm 0.005$
&$+0.035-0.028$ \\
               &      &      &       &  1815     &$\pm 0.004$
&$+0.031-0.025$ \\
$S_3^d$        &--2/3 &--1   &--2.012& 24779     &$\pm 0.009$
&$\pm 0.009$     \\
               &      &      &       & 31493    &$\pm 0.008$
&$\pm 0.008$     \\
\hline
$R_2^u$        &  5/3 &  1/2 &  0.277& 34465    &$\pm 0.005$
&$-0.044+0.041$     \\
               &      &      &       & 44107    &$\pm 0.004$
&$-0.039+0.036$     \\
$R_2^d$        &  2/3 &--1/2 &--1.552& 14793    &$\pm 0.009$
&$\pm 0.009$\\
               &      &      &       & 18815    &$\pm 0.008$
&$\pm 0.008$\\
\hline
$\tilde{R}_2^u$&  2/3 &  1/2 &  0.824&  34465    &$\pm 0.005$
&$\pm 0.042$ \\
               &      &      &       &  44107    &$\pm 0.004$
&$\pm 0.038$ \\
$\tilde{R}_2^d$&--1/3 &--1/2 &--1.006&  9005    &$\pm 0.006$
&$\pm 0.012$ \\
               &      &      &       & 11486     &$\pm 0.005$
&$\pm 0.010$ \\
\hline\hline
\end{tabular}
\end{center}
\renewcommand{\arraystretch}{1}

\vspace{1mm}  \noindent
{\sf Table~1:}~Statistical precision to which the        electro-weak
couplings $Q_\Phi^\gamma$ and $Q_\Phi^Z$ of scalar leptoquarks can
be measured at TESLA. 
First lines: $\sqrt{s} = 800 \GeV, M_\Phi = 320
\GeV, {\cal L} = 500 \fb^{-1}$;
Second lines: $\sqrt{s} = 1 \TeV, M_\Phi = 400
\GeV, {\cal L} = 1 \ab^{-1}$.

\end{document}